\documentclass[prl,twocolumn,superscriptaddress,showpacs,
amsmath,amssymb]{revtex4}

\usepackage{graphicx} 
\usepackage{dcolumn} 

\def\bk{{\bf k}}

\def\bq{{\bf q}}

\def\bv{{\bf v}}

\def\b0{{\bf 0}}

\def\Re{{\rm Re}}
\def\Im{{\rm Im}}
\def\bra{\langle}
\def\ket{\rangle}

\def\eps{\epsilon}

\def\Lam{\Lambda}
\def\om{\omega}

\def\sg{\sigma}
\def\Sg{\Sigma}

\begin{document}

\title{Fermi surface truncation from thermal nematic fluctuations}

\author{Hiroyuki Yamase}
\affiliation{National Institute for Materials Science, 
 Tsukuba 305-0047, Japan}
\author{Walter Metzner}
\affiliation{Max-Planck-Institute for Solid State Research,
 D-70569 Stuttgart, Germany}

\date{\today}

\begin{abstract}
We analyze how thermal fluctuations near a finite temperature
nematic phase transition affect the spectral function $A(\bk,\om)$
for single-electron excitations in a two-dimensional metal.
Perturbation theory yields a splitting of the quasi-particle peak
with a $d$-wave form factor, reminiscent of a pseudogap.
We present a resummation of contributions to all orders in the
Gaussian fluctuation regime.
Instead of a splitting, the resulting spectral function exhibits 
a pronounced broadening of the quasi-particle peak, which varies 
strongly around the Fermi surface and vanishes upon approaching 
the Brillouin zone diagonal.
The Fermi surface obtained from a Brillouin zone plot of $A(\bk,0)$
seems truncated to Fermi arcs.
\end{abstract}

\maketitle


The concept of nematic order in interacting electron liquids has
attracted considerable interest over the last decade, mostly in the
context of (quasi) two-dimensional systems \cite{fradkin10,vojta09}.
In a nematic state an orientational symmetry of the system is
spontaneously broken, without breaking however the translation 
invariance.
One route toward a nematic state is via partial melting of stripe
order in a doped antiferromagnetic Mott insulator \cite{kivelson98}.
Alternatively, a nematic state can be obtained from a Pomeranchuk
\cite{pomeranchuk59} instability generated by forward scattering
interactions in a normal metal \cite{yamase00,halboth00}.
On a square lattice, the most natural candidate for a Pomeranchuk
instability has $d_{x^2-y^2}$ symmetry. 

Signatures of nematic order with a $d_{x^2-y^2}$ symmetry have
been observed in several strongly interacting electron materials.
A nematic phase with a sharply defined phase boundary has been
established for
$\rm Sr_3 Ru_2 O_7$ in a strong magnetic field \cite{ruthenate}.
Nematic order has also been observed in the high temperature
superconductor $\rm Y Ba_2 Cu_3 O_y$ in transport experiments
\cite{daou10} and neutron scattering \cite{hinkov}.
Due to the slight orthorhombicity of the $\rm CuO_2$ planes one 
cannot expect a sharp nematic phase transition in 
$\rm Y Ba_2 Cu_3 O_y$. However, the strong temperature dependence
of the observed in-plane anisotropy indicates that the system 
develops an intrinsic electronic nematicity, which drastically
enhances the in-plane anisotropy imposed by the structure
\cite{hackl09,yamase06}.

Nematic fluctuations close to a continuous nematic quantum phase
transition naturally lead to non-Fermi liquid behavior 
\cite{oganesyan01,metzner03,garst10}.
For a $d$-wave Pomeranchuk instability on a square lattice,
the decay rate of electronic excitations is strongly momentum
dependent along the Fermi surface.
At the quantum critical point, the decay rate for single-particle
excitations is proportional to $d_{\bk}^2 |\om|^{2/3}$, where
$\om$ is the excitation energy and $d_{\bk}$ is a form factor
with $d$-wave symmetry, such as $d_{\bk} = \cos k_x - \cos k_y$ 
\cite{metzner03,metlitski10}.
Landau quasi-particles are thus unstable everywhere on the Fermi
surface except at the Brillouin zone diagonals, where $d_{\bk}$
vanishes.
The temperature dependence of the decay rate in the quantum
critical regime near the quantum critical point also differs
strongly from Fermi liquid behavior \cite{dellanna06,dellanna07}.

Previous works on non-Fermi liquid behavior caused by nematic
fluctuations focussed on the quantum critical point and the 
quantum critical regime at finite temperature.
In this paper we analyze the spectral function for single-particle
excitations in the {\em thermal}\/ fluctuation regime near a 
nematic phase transition at finite temperatures.
We show that a perturbative calculation of the self-energy
in that regime leads to a splitting of the quasi-particle peak
with a $d$-wave form factor in the single-particle excitation 
spectrum, reminiscent of a pseudogap.
However, in a self-consistent calculation the split peak in the
spectral function is replaced by a single broad peak.
In the Gaussian fluctuation regime, a summation of vertex 
corrections to all orders is possible and confirms the 
self-consistent result.
The Fermi surface obtained from the peak of the spectral function
at zero excitation energy is thus smeared by a smooth broadening, 
which is most pronounced near the points $(\pi,0)$ and $(0,\pi)$ of 
the Brillouin zone, while it gradually decreases toward the 
Brillouin zone diagonal.


We consider a one-band system of electrons on a square lattice
with a tight-binding dispersion $\eps_{\bk}$ and an effective 
interaction of the form \cite{metzner03}
\begin{equation} \label{H_I}
 H_I = 
 \frac{1}{2L} \sum_{\bq} g(\bq) \, n_d(\bq) \, n_d(-\bq) \; ,
\end{equation}
where $n_d(\bq) = \sum_{\bk,\sg} d_{\bk} \,
 c_{\bk-\bq/2,\sg}^{\dag} c_{\bk+\bq/2,\sg}$
are $d$-wave density fluctuation operators, and $L$ is the 
number of sites.
The function $g(\bq)$ is negative and peaked at $\bq = 0$, 
so that forward scattering dominates.
An effective interaction of the form $H_I$ can be obtained
from microscopic models such as the Hubbard or $t$-$J$ model 
\cite{yamase00,halboth00}.


For sufficiently negative values of  $g = g(\b0)$ the interaction
generates a $d$-wave Pomeranchuk instability leading to a
nematic state with a spontaneously broken orientation symmetry
\cite{metzner03,kee03,khavkine04,yamase05}.
A suitable order parameter characterizing the symmetry breaking
is provided by the expectation value $\bra n_d(\b0) \ket$.
Close to the transition (if continuous), strong $d$-wave density
fluctuations with a long wavelength develop, which lead to a
singular effective interaction.
In the quantum critical regime the effective interaction is
dynamical and of the form \cite{metzner03,dellanna06}
\begin{equation} \label{D}
 D_{\bk\bk'}(\bq,\nu_n) =
 \frac{g \, d_{\bk} d_{\bk'}}
 {(\xi_0/\xi)^2 + \xi_0^2 |\bq|^2 + |\nu_n|/(u|\bq|)} \; ,
\end{equation}
where $\nu_n = 2\pi n T$ is a bosonic Matsubara frequency; $\xi$ 
is the nematic correlation length, while $\xi_0$ and $u$ are 
non-universal parameters determined by the momentum dependence 
of $g(\bq)$ and the band structure.
In the thermal fluctuation regime near the finite temperature
phase transition, quantum ($\nu_n \neq 0$) fluctuations are cut 
off by temperature, such that only the classical part of the 
effective interaction,
\begin{equation} \label{D_cl}
 D_{\bk\bk'}(\bq) =
 D_{\bk\bk'}(\bq,0) =
 \frac{\tilde g \, d_{\bk} d_{\bk'}}{\xi^{-2} + |\bq|^2} \; ,
\end{equation}
with $\tilde g = g/\xi_0^2$, is important.


The nematic transition on a square lattice belongs to the
two-dimensional Ising universality class.
A thermal phase transition at a critical temperature $T_c > 0$ 
is possible, since the dimensionality of the system is above 
the lower critical dimension (one). 
The correlation length $\xi$ diverges at $T_c$.
Approaching the critical temperature, one first passes
through a Gaussian fluctuation regime, where order parameter
interactions are not important. 
Moving closer to $T_c$, one enters the Ginzburg region, 
where order parameter interactions become relevant, and the 
fluctuation propagator acquires an anomalous scaling 
dimension \cite{onsager44}.
Close to the quantum critical point, the width of the Ginzburg
region is of order $T_c/|\log T_c|$ \cite{bauer11}.


The momentum resolved spectral function for single-particle
excitations can be written as
\begin{equation}
 A(\bk,\om) = - \frac{1}{\pi} \Im G(\bk,\om) =
 - \frac{1}{\pi} 
 \Im \frac{1}{\om - (\eps_{\bk} - \mu) - \Sg(\bk,\om)} \; ,
\label{A}
\end{equation}
where $G(\bk,\om)$ and $\Sg(\bk,\om)$ are the retarded Green 
function and self-energy, respectively.
We first compute the self-energy perturbatively to first order 
in the effective interaction.
The contribution from classical fluctuations is given by 
\cite{dellanna06}
\begin{equation}
 \Sg(\bk,\om) = - T \int \frac{d^2q}{(2\pi)^2} \,
 D_{\bk\bk}(\bq) \, G(\bk-\bq,\om) \; .
\label{Sigma}
\end{equation}

In a non-selfconsistent evaluation of Eq.~(\ref{Sigma}) one 
approximates $G$ by the non-interacting Green function
$G_0(\bk,\om) = [\om - (\eps_{\bk} - \mu) + i0^+]^{-1}$.
The self-energy can then be computed analytically. The imaginary
part has been obtained already previously \cite{dellanna06}. 
For momenta close to the Fermi surface and small frequencies 
one finds
\begin{equation}
 \Im\Sg(\bk,\om) = \frac{\tilde g d_{\bk}^2}{4 v_{\bk}} \,
 T \xi \, l(\kappa) \; ,
\label{ImSigma}
\end{equation}
where $v_{\bk} = |\nabla\eps_{\bk}|$ is the velocity of the
electrons, $\kappa = [\om - (\eps_{\bk} - \mu)] \xi/v_{\bk}$,
and $l(\kappa) = (1 + \kappa^2)^{-1/2}$. 
We assume that the Fermi surface does not cross van Hove points, 
such that $v_{\bk}$ is finite.
The real part of the self energy is obtained either by a 
direct evaluation of Eq.~(\ref{Sigma}) or by a Kramers-Kronig 
transformation of the imaginary part as
\begin{equation}
 \Re\Sg(\bk,\om) = \frac{\tilde g d_{\bk}^2}{4\pi v_{\bk}}
 \, \xi T \, l(\kappa) \, \ln \left| 
 \frac{1 - \kappa l(\kappa)}{1 + \kappa l(\kappa)} 
 \right| \; .
\label{ReSigma}
\end{equation}

In Fig.~1 we show results for the spectral function as 
obtained from the non-selfconsistent first order calculation
of the self-energy.
Here and in all further numerical results we have chosen a
dispersion
$\eps_{\bk} = -2t (\cos k_x + \cos k_y) - 4t' \cos k_x \cos k_y
 - 2t'' (\cos 2k_x + \cos 2k_y)$ 
with hopping amplitudes $t=1$, $t'=-0.3$, and $t''=0.15$.
The lattice constant is one, and the chemical potential $\mu$ 
has been chosen such that the electron density is fixed at 
$n = 0.9$. 
The corresponding Fermi surface is closed around $(\pi,\pi)$.
The coupling constant is $\tilde g = - 1.2$ and the temperature
$T = 0.15$.
We have not attempted to compute the correlation length, 
since it depends on model details such as the momentum
dependence of $g(\bq)$. Instead we show results for various
choices of $\xi$.
One can see that a pronounced splitting of the quasi-particle
peak develops for increasing $\xi$, which could be interpreted 
as a fluctuation precursor of the symmetry-broken state.
However, we now show that the splitting disappears 
in a self-consistent calculation, and it is not restored by 
vertex corrections, at least in the Gaussian fluctuation regime.
\begin{figure}[ht!]
\begin{center}
\includegraphics[width=2.5in]{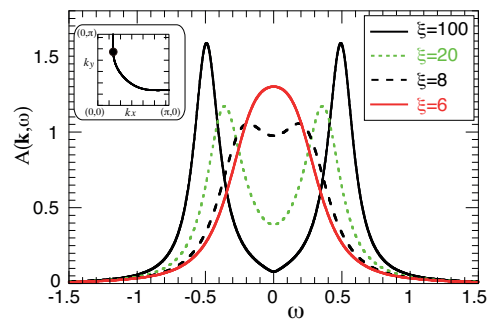}
\caption{(Color online) Spectral function $A(\bk,\om)$ 
 as obtained from a non-selfconsistent perturbative 
 calculation of the self-energy for various choices
 of the correlation length $\xi$.
 The graphs show $A(\bk,\om)$ as a function of $\om$ for 
 a fixed momentum $\bk = (0.69,2.34)$ on the Fermi surface 
 remote from the Brillouin zone diagonal (see inset).}
\end{center}
\end{figure}

In a self-consistent evaluation of Eq.~(\ref{Sigma}), with 
the interacting Green function on the right hand side, one 
has to solve an integral equation.
The problem can be simplified by decomposing the momentum
transfer $\bq$ in components $q_r$ and $q_t$ normal and 
tangential to the Fermi surface, respectively \cite{dellanna06}. 
The dependence on $q_t$ can be neglected in the momentum 
argument of $G$, such that the $q_t$-integral acts only on
the fluctuation propagator $D_{\bk\bk}(\bq)$, yielding
\begin{equation}
 \bar D_{\bk}(q_r) =
 \int \frac{dq_t}{2\pi} D_{\bk\bk}(\bq) = 
 \frac{\tilde g d_{\bk}^2}{2\sqrt{\xi^{-2} + q_r^2}} \; .
\label{Dbar}
\end{equation}
The momentum dependence of the self-energy $\Sg(\bk,\om)$ 
can be parametrized by the Fermi momentum $\bk_F$ closest
to $\bk$ and the oriented distance from the Fermi surface 
$k_r$, which carries the sign of $\xi_{\bk}$.
One then obtains the one-dimensional integral equation
\begin{eqnarray}
 \Sg(k_r,\om) &=& \frac{\tilde g d_{\bk_F}^2 T}{4\pi} 
 \int \frac{dq_r}{\sqrt{\xi^{-2} + q_r^2}} 
 \nonumber \\
 &\times&
 \frac{1}{v_{\bk_F}(k_r - q_r) - \om + \Sg(k_r - q_r,\om)} 
 \; .
\label{selfcons}
\end{eqnarray}
Note that $\Sg$ does not depend on $k_r$ and $\om$ 
independently, but only on the difference $\om - v_{\bk_F}k_r$.
The dependence of $\Sg$ on $\bk_F$ enters only parametrically
via $v_{\bk_F}$ and $d_{\bk_F}$ and has not been written
explicitly.
The integral equation (\ref{selfcons}) can be solved numerically.
The results for $A(\bk,\om)$ differ strongly from those suggested
by non-selfconsistent perturbation theory.
The quasi-particle splitting observed in the perturbative 
calculation (Fig.~1) is wiped out completely by the self-energy 
feedback into $G$.
The spectral function exhibits only a single peak with a maximum
at $\om = v_{\bk_F} k_r$, even for a very large correlation
length $\xi$.

We now consider higher order contributions not contained in the 
self-consistent one-loop approximation (\ref{Sigma}).
The sum over all self-energy contributions generated by 
thermal fluctuations can be written as
\begin{eqnarray}
 \Sg(\bk,\om) &=& - T \int \frac{d^2q}{(2\pi)^2} \,
 D_{\bk\bk}(\bq) \, G(\bk-\bq,\om)
 \nonumber \\
 &\times& \Lam(\bk-\bq/2,\om;\bq,0) \; ,
\label{Sg_exact}
\end{eqnarray}
where $\Lam(\bk,\om;\bq,\nu)$ is the irreducible charge vertex 
including all vertex corrections.
We exploit the fact that dominant contributions are due 
to small momentum transfers $\bq$ of order $\xi^{-1}$, 
to resum vertex corrections via an asymptotic Ward identity 
\cite{castellani94,metzner98}.
For small $\bq$, the charge vertex is related to the current
vertex ${\bf\Lam}$ and the propagator via the Ward identity
$\nu \Lam(\bk,\om;\bq,\nu) - \bq \cdot {\bf\Lam}(\bk,\om;\bq,\nu)
 = G^{-1}(\bk+\bq/2,\om+\nu/2) - G^{-1}(\bk-\bq/2,\om-\nu/2)$.
In the Gaussian fluctuation regime, interactions between order 
parameter fluctuations are not important. In a diagrammatic
representation of perturbation theory, these interactions are
generated by fermionic loops with more than two vertices.
Neglecting Feynman diagrams with such loops leads to two
simplifications. First, the effective interaction (\ref{D_cl})
remains unrenormalized. Second, diagrams contributing to the
charge and current vertices involve only an open fermionic
line. Since contributions with small $\bq$ dominate, the 
electron velocity $\bv_{\bk}$ entering the current operator
is almost conserved such that the current vertex can be 
expressed by the charge vertex as 
${\bf\Lam}(\bk,\om;\bq,\nu) = \bv_{\bk} \Lam(\bk,\om;\bq,\nu)$.
The latter relation holds for each Feynman diagram without 
fermionic loops.
Combining this with the Ward identity one obtains, in the 
static limit $\nu = 0$,
\begin{equation}
 \Lam(\bk,\om;\bq,0) =
 \frac{ G^{-1}(\bk-\bq/2,\om) - G^{-1}(\bk+\bq/2,\om)}
 {\bv_{\bk} \cdot \bq} \; .
\label{ward}
\end{equation}
Inserting Eq.~(\ref{ward}) into Eq.~(\ref{Sg_exact}), and using the 
Dyson equation $G^{-1} = G_0^{-1} - \Sg$, one obtains a closed 
system of equations for $\Sg$ and $G$. 
Decomposing the momenta $\bk$ and $\bq$ in radial and tangential
components, and integrating $D_{\bk\bk}(\bq)$ over $q_t$ as before
(self-consistent solution), one finds a one-dimensional linear 
integral equation for $G$,
\begin{eqnarray}
 && (\om - v_{\bk_F} k_r + i0^+) \, G(k_r,\om) = \nonumber \\ 
 && \hskip 1cm 1 + 
 T \int \frac{dq_r}{2\pi} \frac{\bar D_{\bk_F}(q_r)}{v_{\bk_F} q_r} 
 \, G(k_r - q_r,\om) \; ,
\label{G_exact}
\end{eqnarray}
with $\bar D_{\bk_F}(q_r)$ from Eq.~(\ref{Dbar}).

Note that vertex corrections cannot be summed by the above method 
at a nematic quantum critical point or for the related problem
of non-relativistic fermions coupled to a $U(1)$ gauge field.
This is because in these cases the dominant momentum transfers
are almost tangential to the Fermi surface, so that the term
$\bv_{\bk} \cdot \bq$ becomes subleading compared to contributions
originating from fluctuations of the electron velocity
\cite{metzner98}.

The integral equation (\ref{G_exact}) can be converted to a
linear differential equation by a Fourier transformation.
The differential equation can be solved by standard methods.
The result for the spectral function reads
\begin{equation}
 A(\bk,\om) = \int_{-\infty}^{\infty} dx \, \hat A(x) \,
 e^{i(\om - v_{\bk_F} k_r)x/v_{\bk_F}} \; ,
\label{A_exact}
\end{equation}
where
\begin{equation}
 \hat A(x) = \frac{1}{2\pi v_{\bk_F}} \,
 \exp \left[ \int_0^x dx' \int_0^{x'} dx'' \,
 T \frac{\tilde g d_{\bk_F}^2}{2\pi v_{\bk_F}^2} \, K_0(x''/\xi)
 \right] \, .
\label{Ax}
\end{equation}
$K_0$ is a modified Bessel function. Note that $A(\bk,\om)$ depends
on $k_r$ and $\om$ only via the difference $\om - v_{\bk_F} k_r$.

In Fig.~2 we plot the spectral function $A(\bk,\om)$ for the same
parameters as in Fig.~1. The function exhibits only a single peak
with no trace of a splitting. The splitting present in Fig.~1 is
therefore an artefact of the perturbation expansion, at least in 
the Gaussian regime.
Vertex corrections do not change the self-consistent one-loop
result qualitatively. Quantitatively they tend to sharpen the peak
in $A(\bk,\om)$, but only moderately.
For a large correlation length $\xi$ the width of the peak in
$A(\bk,\om)$ is proportional to $\sqrt{\log\xi}$, corresponding 
to a peak in the imaginary part of the self-energy
$\Im\Sg(\bk,\om) \propto \sqrt{\log\xi}$ at $\om = v_{\bk_F} k_r$.
The width of the peak in $\Im\Sg$ is also proportional to 
$\sqrt{\log\xi}$ and therefore increases with $\xi$.
This is very different from the perturbative result for $\Im\Sg$, 
Eq.~(\ref{ImSigma}), where the height of the peak increases
rapidly with $\xi$, while its width shrinks.
\begin{figure}[ht!]
\begin{center}
\includegraphics[width=2.5in]{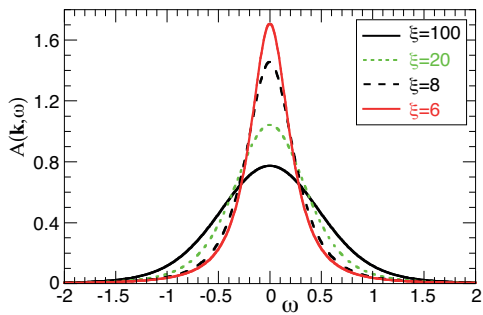}
\caption{(Color online) Spectral function $A(\bk,\om)$
 as obtained from a non-perturbative resummation of 
 contributions from thermal fluctuations, including
 vertex corrections.
 The choice of $\bk$ and the model parameters are the 
 same as in Fig.~1.}
\end{center}
\end{figure}

In the quantum critical regime studied previously \cite{dellanna06}
the spectral function also exhibits a single peak with a temperature
dependent broadening. In that regime the width of the peak is
proportional to $T\xi$, with a correlation length $\xi$ diverging
as $(T |\log T|)^{-1/2}$ upon approaching the quantum critical
point at $T = 0$.

It is striking that perturbation theory indicates a fluctuation
precursor of the symmetry broken state at leading order, which is 
however not robust with respect to higher order contributions.
One may compare with the case of a charge density wave with a 
finite wave vector, where symmetry breaking opens a gap.
Perturbation theory indicates a pseudogap above the transition 
temperature in such systems, for example for the flux order 
studied in Ref.~\cite{greco09}.
It would be interesting to analyze the fate of the pseudogap
in such systems in a calculation beyond perturbation theory.

Due to the $d$-wave form factor in the effective interaction,
the broadening of the spectral function varies strongly in 
momentum space. To illustrate this, we plot an intensity map
of $A(\bk,0)$ in the first quarter of the Brillouin zone in
Fig.~3.
$A(\bk,0)$ as obtained from Eq.~(\ref{A_exact}) diverges on
the Brillouin zone diagonal, since $d_{\bk}$ vanishes there.
In a more complete model, where other interaction channels 
should be added, this divergence will be cut off at least 
by regular (Fermi liquid) contributions to $\Im\Sg$ of 
order $T^2$. We have therefore included such a regular 
contribution ($-T^2$).
Due to the rapid increase of the broadening of the peak in
$A(\bk,0)$ upon moving away from the Brillouin zone diagonal,
and the corresponding decrease in the peak height, the
Fermi surface seems truncated to arcs, albeit with fuzzy
ends.
\begin{figure}[ht!]
\begin{center}
\includegraphics[width=1.8in]{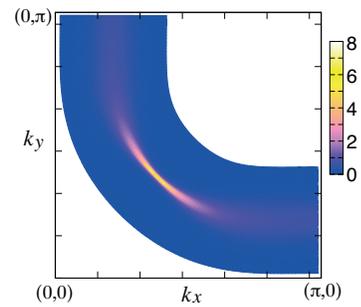}
\caption{(Color online)
 Intensity plot of the spectral function $A(\bk,0)$ as a 
 function of $\bk$ in the first quarter of the Brillouin 
 zone at $T = 0.2$, for a correlation length $\xi = 15$. 
 The other parameters are the same as in Figs.~1 and 2.}
\end{center}
\end{figure}
%


In summary, we have computed the spectral function $A(\bk,\om)$
for single electron excitations in the presence of critical 
fluctuations near a thermal nematic phase transition in a 
two-dimensional metal.
Leading order perturbation theory indicates a splitting of the 
quasi-particle peak.
However, a resummation of contributions to all orders reveals
that the splitting is an artefact of perturbation theory, at
least in the Gaussian fluctuation regime.
The spectral function exhibits a pronounced broadening with 
a $d$-wave form factor, leading to features reminiscent of 
Fermi arcs in the Brillouin zone.
The qualitative shape of the spectral function does not depend 
on the specific choice of parameters.
Away from the Brillouin zone diagonal, the imaginary part of the 
self-energy has a peak at $\om = \eps_{\bk} - \mu$, in clear contrast 
to the conventional Fermi liquid form.

It is remarkable that the effect of Gaussian thermal fluctuations 
on electronic excitations could be treated non-perturbatively.
The method used to sum contributions from thermal fluctuations to
all orders is not restricted to nematic fluctuations, but could
be applied equally well to other critical thermal fluctuations 
with a small wave vector, for example, close to a structural
phase transition.

A continuous finite temperature phase transition is well
established at the roof of the nematic dome found for 
$\rm Sr_3 Ru_2 O_7$ in a strong magnetic field \cite{ruthenate}.
That system thus provides an opportunity to observe the fluctuation 
effects computed in this work, but also an experimental challenge, 
since the standard tool to measure the momentum resolved spectral 
function, that is, photoemission, is hampered by the magnetic 
field.

Fermi arcs have been observed in photoemission measurements of
the spectral function $A(\bk,\om)$ in various high-$T_c$ cuprate
compounds \cite{damascelli03}.
However, the Fermi surface truncation in these materials is
associated with a pseudogap formation, while we obtain only a
strongly momentum dependent broadening of the spectral function.
Although experiments indicate electronic nematicity at least 
in some cuprates \cite{fradkin10}, our results thus show that 
another mechanism needs to be invoked to explain the photoemission 
data.

\vspace*{5mm}

\begin{acknowledgments}
We are grateful to A.~Chubukov, A.~Greco, A.~Katanin, B.~Obert,
and A.~Rosch for valuable discussions, and to J.~Bauer for a 
critical reading of the manuscript.
\end{acknowledgments}


\end{document}